# Fabrication and transport critical currents of multifilamentary $MgB_2$/Fe wires and tapes

H.L. Suo, C. Beneduce, X.D. Su and R. Flükiger

Département de Physique de la Matière Condensée, Université de Genève, 24 quai Ernest-Ansermet, CH-1211 Genève 4, Switzerland

*Abstract* Multifilamentary $MgB_2$/Fe wires and tapes with high transport critical current densities have been fabricated using a straightforward powder-in-tube (PIT) process. After annealing, we measured transport $j_c$ values up to $1.1 \times 10^5$ A/cm$^2$ at 4.2 K and in a field of 2 T in a $MgB_2$/Fe square wire with 7 filaments fabricated by two-axial rolling, and up to $5 \times 10^4$ A/cm$^2$ at 4.2 K in 1 T in a $MgB_2$/Fe tape with 7 filaments. For higher currents these multifilamentary wires and tapes quenched due to insufficient thermal stability of filaments. Both the processing routes and deformation methods were found to be important factors for fabricating multifilamentary $MgB_2$ wires and tapes with high transport $j_c$ values.

INTRODUCTION

The recent discovery [1] of $MgB_2$ as a superconductor with transition temperature near 40 K has promoted considerable interest worldwide in the area of basic and applied research on superconducting materials. High transport critical current densities have been reported in $MgB_2$ bulk samples [2]-[4]. Due to its intermediate transition temperature, weak-link free grain boundaries [5] and low material cost, this material is interesting for potential applications such as MRI (magnetic resonance imaging) and transformers at operating temperature of 20-30K. Soon after its discovery, several research groups have tried to demonstrate the feasibility of fabrication of $MgB_2$ wires or tapes with high critical current density using the Powder-In-Tube (PIT) method [6-11]. Large critical current densities of $10^6 A/cm^2$ and significant enhancement of irreversibility line and $H_{c2}$ have been reported in $MgB_2$ films grown on single-crystalline substrates [12-13]. These results brought further excitement to the development of $MgB_2$ in view of an eventual industrial application.

We have previously reported [6, 14] on the preparation of highly dense monofilamentary $MgB_2$ tapes with large transport critical current density using Ni and Fe sheath materials. It was found that both grain size of starting powder and appropriate deformation and heat treatment were determinant for getting higher $j_c$ values. The transport $j_c$ values in annealed monofilamentary $MgB_2$/Fe tapes were up to $10^4$ $A/cm^2$ at 4.2 K and 6.5 T and the extrapolated self-field value of $j_c$ in this tape was close to $10^6$ $A/cm^2$ [6, 14]. Fe was found to be an excellent candidate material for preparation of $MgB_2$ tapes or wires. So far, it was reported that multifilamentay $MgB_2$ wires were produced either using Cu-Ni tube [15] or Cu/NbZr sheath material [16]. In this paper, we extend the use of Fe sheaths [14] to multifilamentary $MgB_2$ wires and tapes. We found a significant effect of the deformation method on the transport $j_c$ values. The highest $j_c$ values in $MgB_2$/Fe wires with 7 filaments were achieved by two-axial rolling, followed by a heat treatment. The effect of both preparation route and deformation technique on fabrication and superconducting properties of multifilamentary wires and tapes is discussed.

EXPERIMENTAL

Two experimental routes were used to fabricate $MgB_2$/Fe wires and tapes with 7 filaments. A schematic diagram of these two routes is shown in Figure 1. The operation process of filling powder was always carried out inside a glove box under an Ar atmosphere. The powder used was a standard commercial $MgB_2$ powder from Alfa-Aesar with a purity of 98% which was ball milled for 2 hours down to smaller grain sizes. Figure 2 shows the distributions of powder sizes with and without ball milling. The as–purchased powder contains a large number of agglomerated grains, with a wide size distribution centred at around 60 µm. After 2 hours of ball milling, 35% of grains have diameters around 3 µm, while 60% are centred around 30 µm. The effect of the powder size on $j_c$ was already presented in our previous work [14].

*A: Route 1 (bundling)*

The first route consisted in packing 7 thin $MgB_2$/Fe single wires into a Fe tube (OD: 10mm, ID: 7mm) and repeating the same deformation procedure (swaging and drawing used for preparation of single wires) up to a wire diameter of 2.9 mm. The drawn wire was deformed either by groove rolling to 1.7 mm multifilamentary wire and two-axial rolling to square wires with diameter 1.7 mm or 1.1 mm, or by flat rolling to a multifilamentary tape with a thickness of 0.9 mm.

*B: Route 2 (drilled holes)*

In order to get thinner $MgB_2$ multifilamentary tapes with well distributed filaments, a second route was followed, consisting in drilling 7 holes with a diameter of 3 mm in a 19 mm Fe rod and packing $MgB_2$ powder in each hole, followed by swaging and drawing up to 5.5 mm. Part of drawn wire was directly deformed by two-axial rolling down to a square wire with diameter of 1.7 mm. A recovery annealing at 800° for 5 hours was performed for softening the Fe sheath which showed considerably work hardening during wire drawing. The annealed wire was drawn again and then deformed either by two-axial rolling to square wires with diameter 1.3 mm or by flat rolling to tapes with a thickness of 0.5 mm. For monofilamentary $MgB_2$/Fe tapes, it has been reported [14] that annealing strongly improves the intergranular connectivity and hence leads to higher critical currents. Therefore all wires and tapes prepared by both route1 and route 2 were finally annealed at 950°C for 0.5 h in a flowing Ar atmosphere.

*C: Sample characterization*

The cross sections and microstructure of all wires and tapes were observed using a Leo-438VP scanning electron microscope equipped with a Kevex system which allows EDX microanalysis. Transport critical currents were measured on 45 mm long wires or tapes pieces in a He bath at 4.2 K by standard four-probe method. The current and voltage contacts were directly soldered to the sheath materials of the samples and the distance between the voltage contacts was approximately 10 mm, and the voltage criterion used was 1 µV/cm. The magnetic field was parallel to the samples surface and perpendicular to the current direction.

RESULTS

*A: Multifilamentary wires and tapes fabricated using route 1 (bundling)*

*1) Effect of deformation method on transport $j_c$*

Figure 3 shows transversal cross-sections of polished $MgB_2$/Fe wires and tape with 7 filaments fabricated using the route 1. After groove rolling and two-axial rolling, the multifilamentary wires have uniform transversal sections, as can be seen from Fig. 3.a and 3.b, respectively. However, after flat rolling, the tape does not present well-distributed $MgB_2$ filaments, as shown in Fig. 3.c. Particularly, the $MgB_2$ core in the middle of the tape is distorted, due to the gliding and cutting of neighboring Fe tubes during deformation. The thinner the tapes, the worse the distribution of $MgB_2$ cores has been found.

Fig. 4 shows the transport $j_c$ values at 4.2 K as function of applied field in annealed $MgB_2$/Fe wires with 7 filaments. The wire produced by groove rolling yields a transport $j_c$ value of $9 \times 10^4 A/cm^2$ at 4.2K and 0.75T. In a field of 2.25T, this wire has a $j_c$ value of $1.7 \times 10^4 A/cm^2$, i.e. 3 times less compared to $5 \times 10^4 A/cm^2$ (4.2K, 2.25T) for the 1.7 mm square wire prepared by two-axial rolling. The $j_c$ data of annealed 7 filaments tape produced by route 1 are not shown in the figure due to its low transport $j_c$ value (around $10^2 A/cm^2$ at 4.2K and 0T). Different deformation processes lead to different $j_c$ values in multifilamentary wires and tapes, thus indicates that deformation is an important factor for getting multifilamentary $MgB_2$ wires with higher $j_c$ values. Groove rolling is known to have multi-axial compression components, which lead to a higher $MgB_2$ core density than for the wire after drawing [17], but this method also could induce some micro-cracks in the filaments due to its fast deformation speeds. Our two-axial rolling device works at slower rolling speed, which leads not only to a higher density, but also to less cracks in the $MgB_2$ core. This explanation can be confirmed by the SEM micrographs of both polished $MgB_2$ cores as shown in figure 5 (a) and (b). A rough and porous microstructure of MgB$_2$ core is observed in wire produced by groove rolling. The high porosity and poor microstructure in the MgB$_2$ core of this wire is likely due to the fast deformation speed of our machine, leading to a degradation of the transport $j_c$ value in this wire, as mentioned above. This is in contrast to the relatively flat and dense aspect of MgB$_2$ cores of square wires made by two-axial rolling. This important improvement of the MgB$_2$ core quality is believed to be responsible for the higher transport $j_c$ in the square wire.

*2) Effect of thickness on transport $j_c$*

A transport $j_c$ value for the 1.1 mm square wire prepared by two-axial rolling is also shown in Figure 4. We measured $j_c$ value of $1.1\times10^5$ A/cm$^2$ (4.2K, 2T) in this wire, the extrapolated value at 4.2K, 0T is around $4\times10^5$ A/cm$^2$. Comparing the field behaviour of both 1.7 and 1.1 mm square wires fabricated by two-axial rolling, we observe similar $j_c$ values (Figure 4) with a cross aspect at 4.5T in both wires, which indicates that the transport $j_c$ values in thinner square wires are comparable with those of the thicker square wires (1.7 mm). The result shows that the effect of thickness on $j_c$ in annealed wires is almost negative.

*B: Multifilamentary wires and tapes fabricated using route 2 (drilled holes)*

*1) Effect of processing route on preparation and transport $j_c$ of multifilamentary wires and tapes*

Due to the observed gliding and cutting of neighboring Fe tubes during deformation at the conditions chosen in the present work, it is difficult to prepare MgB$_2$ multifilamentary tape with a well-distributed MgB$_2$ cores using route 1 (bundling). The second route was thus designed to solve these problems. Figure 6 shows the transversal cross-section of polished 0.5 mm thick MgB$_2$/Fe tape with 7 filaments fabricated using route 2 (a Fe rod with drilled holes). Compared with Figure 3 (c), the multifilamentary tape produced by route 2 has a very uniform transversal section and homogeneous distribution of MgB$_2$ filaments. Figure 7 illustrates the transport $j_c$ values at 4.2 K as function of applied field in annealed MgB$_2$/Fe tape and square wires fabricated by route 2. Firstly, the tape produced by flat rolling in route 2 yields a transport $j_c$ value of $5\times10^4$ A/cm$^2$ at 4.2K and 1T, i.e. more than one order of magnitude larger than for the tape prepared by route 1. In the MgB$_2$ square wire produced by route 2, we obtained a transport critical current density of $4.8 \times 10^4$ A/cm$^2$ at 4.2K and 3.5T, which is comparable with that of the square wire produced using route 1 (shown in the Figure 4). The self-field $j_c$ value was not accessible in the present square wire due to the insufficient thermal stability as a consequence of the high electrical resistivity of the Fe sheath. Extrapolating the field dependence of transport $j_c$ in this wire yields self-field values at 4.2K around $4 \times 10^5$ A/cm$^2$, which is somewhat smaller than the values reported in our monofilamentary MgB$_2$/Fe tapes [6]. This may be due to a higher degree of texturing in the monofilamentary MgB$_2$/Fe tapes [18]. Further enhancement of $j_c$ in multifilamentary wires and tapes is expected after improving the texture in the MgB$_2$ filaments. Compared with those of MgB$_2$/Fe square wires with 7 filaments, the lower $j_c$ values in the tape with 7 filaments might be due to the effect of intermediate annealing which is discussed below. The present work demonstrates a significant improvement of the $j_c$ values in multifilamentary tape prepared by route 2 (drilled).

*2) Effect of intermediate annealing on transport $j_c$*

As shown in Figure 7, the transport $j_c$ value in the thinner square wire (1.3 mm) fabricated by applying intermediate annealing during deformation is slightly lower compared with that of the thicker wire (1.7 mm) produced without intermediate annealing. This lowering of the transport $j_c$ value may be due to the shrinkage of the MgB$_2$ cores occurring during the intermediate annealing, resulting in cracks in the cores when performing further deformation. The tape produced by flat rolling is also employed an intermediate annealing at 800°C during deformation. The lower $j_c$ in this tape is also an evidence that intermediate annealing might be one of reason for the degradation of the $j_c$ value of the MgB$_2$ wires or tapes. The further optimizations and effect of intermediate annealing and deformation process are being investigated.

CONCLUSION

In summary, we report on the fabrication and superconducting properties of multifilamentary MgB$_2$ tapes and wires using Fe sheaths. For MgB$_2$/Fe square wires with 7 filaments, we found that the highest $j_c$ values are obtained after two-axial rolling, followed by annealing. The $j_c$ value in these MgB$_2$/Fe square wires after annealing at 950°C was $1.1\times10^5$ A/cm$^2$ at 4.2 K and 2 T. The estimated self-field value of $j_c$ at 4.2 K in these multifilamentary wires was close to $4\times10^5$ A/cm$^2$. We found a significant variation of the $j_c$ values using different deformation methods. Compared with groove rolling, two-axial rolling increased the cores density, thus raising $j_c$. The result indicated that there was no effect of the wire dimensions on the transport $j_c$ values. A 7-filament MgB$_2$ tape with a transport $j_c$ value of $5\times10^4$ A/cm$^2$ at 4.2K and 1T was also obtained. The preparation processing was found significant for getting multifilamentary MgB$_2$ tapes with high $j_c$ values.

ACKNOWLEDGMENT

We thank Patrick Cerutti and Aldo Naula for their technical support in the tape annealing. Dr. Pierre Toulemonde, Dr. Reynald Passerini, Gregoire Witz and Paola Lezza for useful discussions and helps. This work was supported by the Fond National Suisse de la Recherche Scientifique.

FIGURE CAPTIONS

**FIGURE. 1.** Schematic diagram of processing route used for fabrication of MgB$_2$/Fe multifilamentary wires and tapes; (a) route 1; (b) route 2.

**FIGURE. 2.** Distribution of powder grain sizes: (a) the as-purchased powder; (b) the ball-milled powder.

**FIGURE. 3** Transversal cross-sections of MgB$_2$/Fe wires and tape with 7 filaments fabricated using route 1 (bundling): (a) wire after groove rolling (diameter: 1.7 mm); (b) square wire after two-axial rolling (dimension: 1.7 mm); (c) tape after flat rolling (thickness: 0.9 mm).

**FIGURE. 4.** Transport critical current densities at T = 4.2 K as function of applied field in annealed 7-filament MgB$_2$/Fe wires prepared by the bundling.

**FIGURE. 5**. SEM images of MgB$_2$ cores in MgB$_2$/Fe wires using route 1(bundling): (a) wire after groove rolling (diameter: 1.7 mm); (b) square wire after two-axial rolling (dimension: 1.7 mm)

**FIGURE. 6**. Transversal cross-sections of MgB$_2$/Fe tapes with 7 filaments using route 2 (drilled holes): tape after flat rolling (thickness: 0.5 mm).

**FIGURE. 7.** Transport critical current densities at T = 4.2 K as function of applied field in annealed 7-filament $MgB_2$/Fe wires and tape using route 2 (drilled holes). For comparison, the transport $j_c$ curve of an annealed monofilamentary $MgB_2$/Fe tape is shown [6]

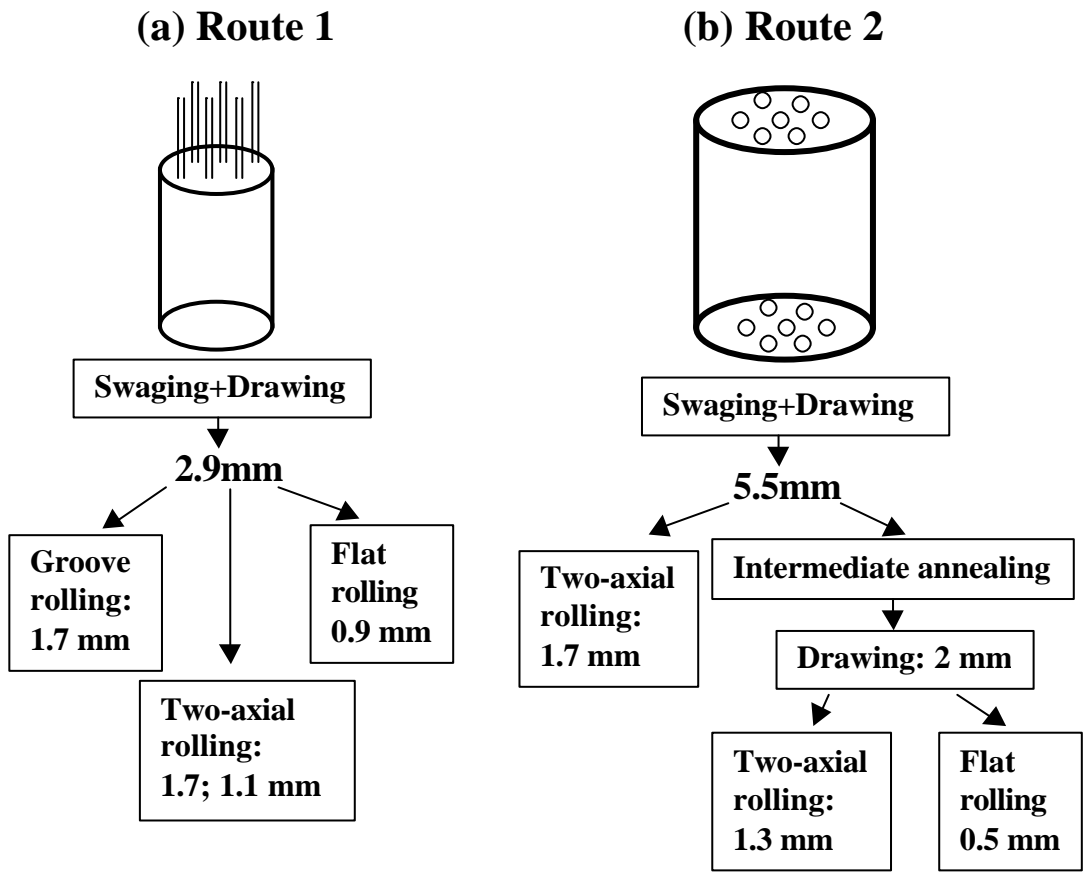

**FIGURE. 1**

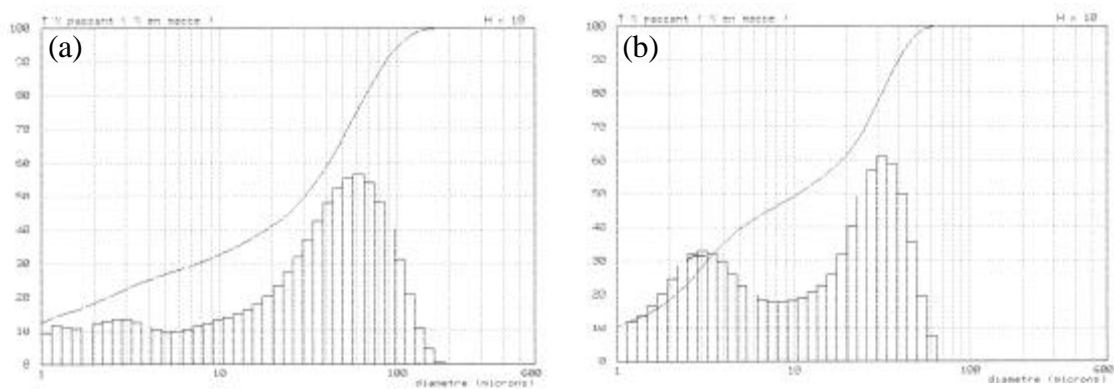

**FIGURE. 2**

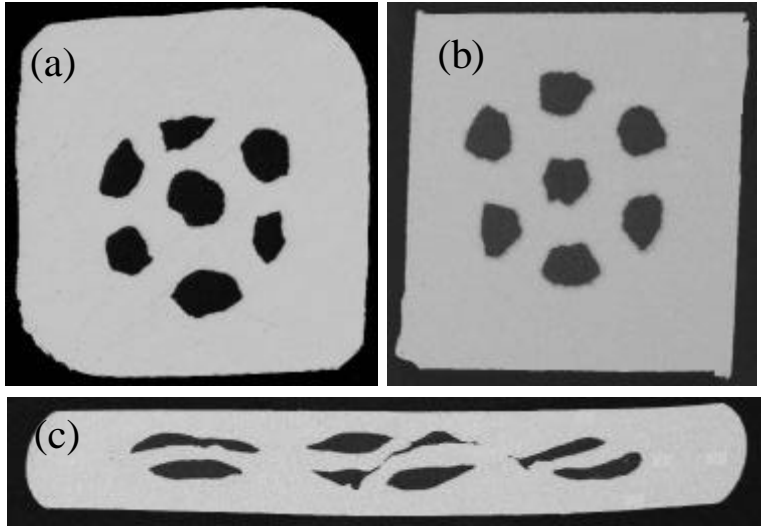

**FIGURE. 3**

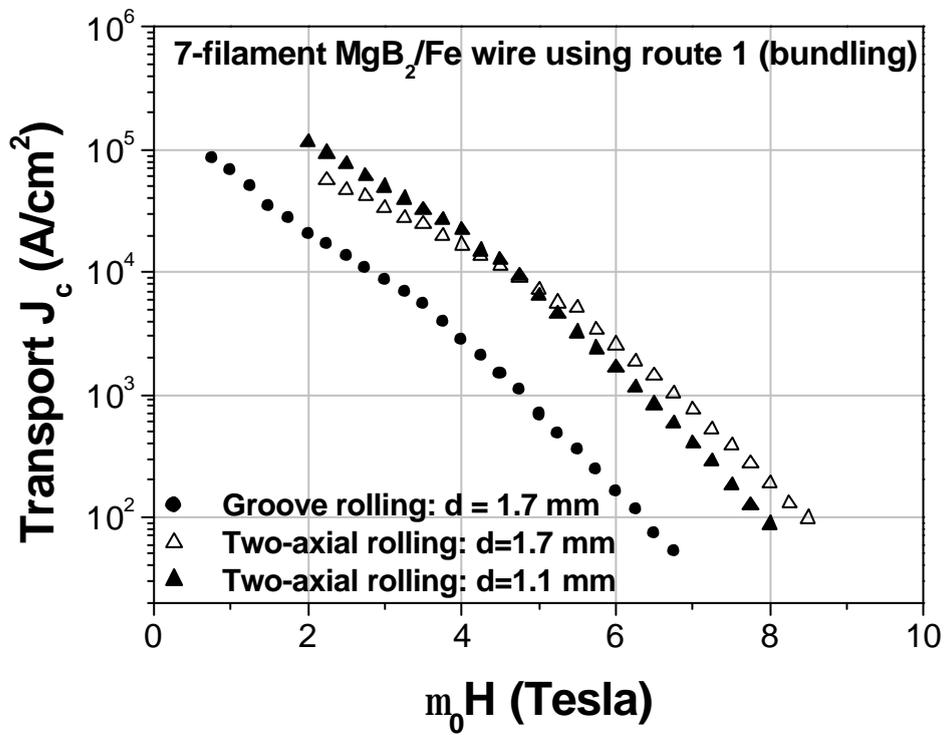

**FIGURE. 4**

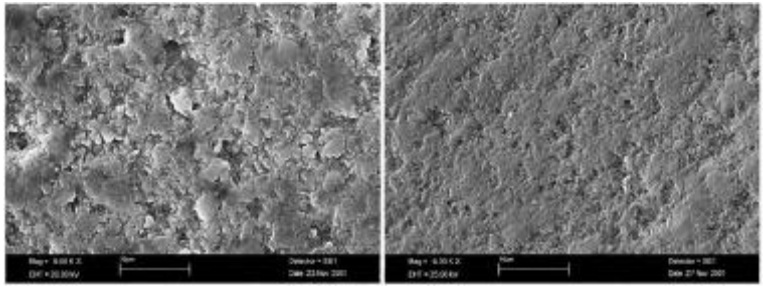

(a)                        (b)

**FIGURE. 5**

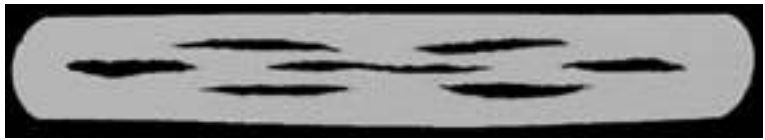

**FIGURE. 6**

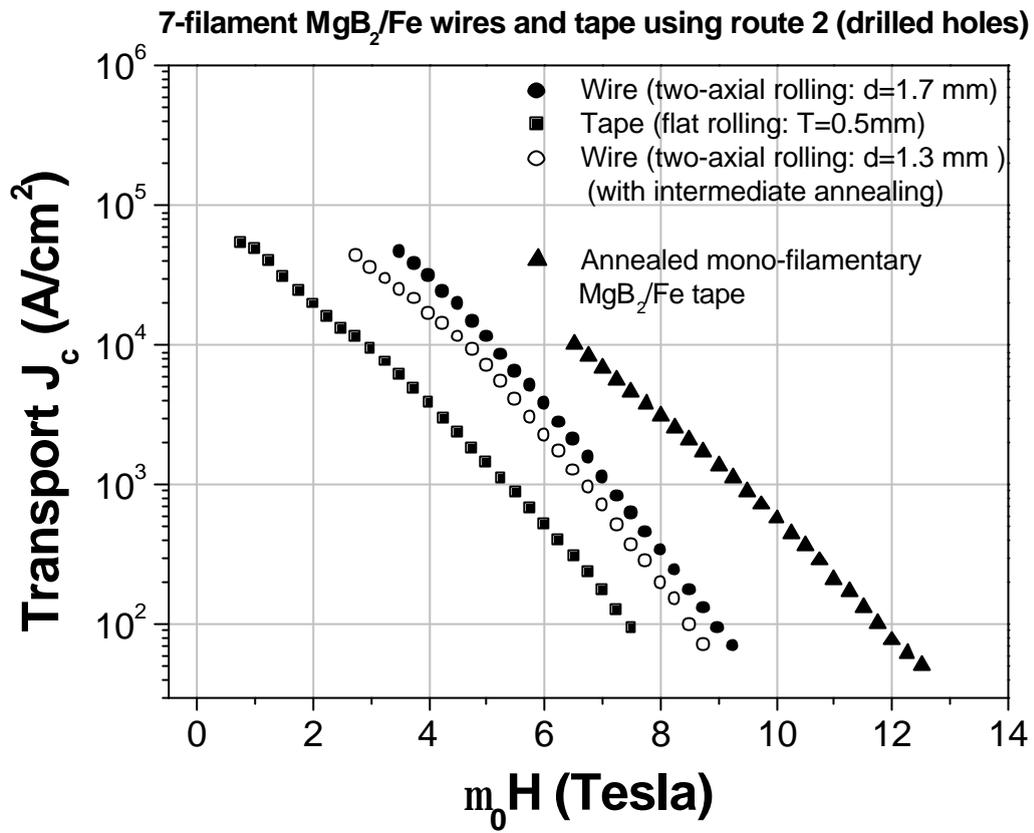

**FIGURE. 7**